\documentclass[%
 aip,
 amsmath,amssymb,
reprint, onecolumn 
]{revtex4-1}

\usepackage{graphicx}
\usepackage{dcolumn}
\usepackage{bm}

\usepackage[utf8]{inputenc}
\usepackage[T1]{fontenc}
\usepackage{mathptmx}
\usepackage{etoolbox}

\makeatletter
\def\@email#1#2{%
 \endgroup
 \patchcmd{\titleblock@produce}
  {\frontmatter@RRAPformat}
  {\frontmatter@RRAPformat{\produce@RRAP{*#1\href{mailto:#2}{#2}}}\frontmatter@RRAPformat}
  {}{}
}%
\makeatother
\begin{document}

\preprint{AIP/123-QED}

\title[H{\"u}ckel theory for M{\"o}bius Nanobelts]{Simple H{\"u}ckel Molecular Orbital Theory for M{\"o}bius Carbon Nanobelts}
\author{Yang Wang}%
 \email{yangwang@yzu.edu.cn}
\affiliation{ 
School of Chemistry and Chemical Engineering, Yangzhou University, Yangzhou, Jiangsu 225002, China 
}%

\date{\today}

\begin{abstract}
The recently synthesized M{\"o}bius carbon nanobelts (CNBs) have gained attention 
owing to their unique $\pi$-conjugation topology, which results in distinctive electronic properties 
with both fundamental and practical implications.
Although M{\"o}bius conjugation with phase inversion in atomic orbital (AO) basis 
is well-established for monocyclic systems, 
the extension of this understanding to double-stranded M{\"o}bius CNBs remains uncertain. 
This study thoroughly examines the simple H{\"u}ckel molecular orbital (SHMO) theory 
for describing the $\pi$ electronic structures of M{\"o}bius CNBs. 
We demonstrate that 
the adjacency matrix for M{\"o}bius CNB 
can preserve its eigenvalues and eigenvectors (with possibly flipped directions) 
under different placements of the sign inversion, 
ensuring identical SHMO results regardless of AO phase inversion location. 
Representative examples of M{\"o}bius CNBs, including the experimentally synthesized one, show that 
the H{\"u}ckel molecular orbitals (MOs) strikingly resemble the DFT-computed $\pi$ MOs, 
which were obtained using a herein proposed technique based on the localization and re-delocalization of DFT canonical MOs. 
Interestingly, the lower-lying $\pi$ MOs exhibit an odd number of nodal planes and are doubly quasidegenerate 
as a consequence of the phase inversion in M{\"o}bius macrocycles, 
contrasting with macrocyclic H{\"u}ckel systems. 
Coulson bond orders derived from SHMO theory correlate well  
with DFT-calculated Wiberg bond indices for all C--C bonds in tested M{\"o}bius CNBs. 
Additionally, a remarkable correlation is observed between HOMO--LUMO gaps 
obtained from the SHMO and GFN2-xTB calculations for a large number of topoisomers of M{\"o}bius CNBs. 
Thus, the SHMO model not only captures the essence of $\pi$ electronic structure of M{\"o}bius CNBs, 
but also provides reliable quantitative predictions comparable to DFT results.  
\end{abstract}

\maketitle

\section{\label{sec:intro}
Introduction
}

Molecules with a $\pi$-conjugation of M{\"o}bius topology have long attracted attention  
because of their peculiar structural and electronic properties 
with significant implications for both fundamental research and practical applications.
\cite{Rzepa:2005aa,Herges:2006aa,Yoon:2009aa,Schaller:2013aa,Sola:2022aa} 
In a M{\"o}bius $\pi$-conjugated system, 
the constituting atoms form a twisted, looped surface that resembles a M{\"o}bius strip, \cite{Richeson:2008}
which has only one continuous side. 
As a consequence, the alignment of the atomic orbitals (AOs) participating in the $\pi$-conjugation 
undergoes an inevitable phase inversion at some point along the cyclic path of the conjugated system. 
By contrast, no such phase shift in AO basis occurs in a H{\"u}ckel $\pi$-conjugated system, 
which is the predominant case in conjugated molecules. 
The unusual M{\"o}bius topology, as compared to the H{\"u}ckel system, 
can result in different electronic structures and hence distinct physical and chemical properties.

A well-known example in this regard is the M{\"o}bius aromaticity \cite{Sola:2022aa,Sola:2022ab,Agranat:2024aa} 
for monocyclic $\pi$-conjugated systems. 
Due to the phase inversion in AO basis, 
a M{\"o}bius conjugated monocycle follows the $4n$ electron-counting rule, 
in lieu of the well-established $4n+2$ rule \cite{Huckel70,Huckel72,Huckel76,E.-Doering:1951ux,Roberts:1952tb} 
for monocyclic H{\"u}ckel systems. 
In other words, a monocyclic conjugated system with M{\"o}bius topology 
is aromatic if the number of $\pi$ electrons is a multiple of four. 
This rule has been corroborated by direct computational studies of aromaticity 
and by the successful synthesis of some intriguing molecules with a M{\"o}bius monocycle 
containing $4n$ delocalized $\pi$ electrons. 
M{\"o}bius-aromatic molecules can thus far be classified into two categories, 
metallacyclic and macrocyclic species,\cite{Sola:2022aa} 
possessing the Craig\cite{CRAIG:1958aa,Craig:1959aa} and Heilbronner types of M{\"o}bius aromaticity, respectively.
The Craig-type M{\"o}bius aromaticity is achieved in metallacyclic compounds 
by introducing to the $\pi$-conjugation  
a d orbital (from a transition metal or a main group element like S and P), 
in which the sign of basis orbitals is inherently inverted. 
The experimentally synthesized Craig-aromatic compounds are exemplified by 
osmabenzene \cite{Elliott:1982aa} and ReB$_4^{0/-1}$ atomic clusters.\cite{Cheung:2020ab}
On the other hand, the Heilbronner-type M{\"o}bius aromaticity arises 
in macrocyclic molecules where the phase shift in AOs is realized 
by structurally twisting the conjugated framework into a M{\"o}bius form.  
Typical examples include successfully synthesized macrocycles of  
annulenes \cite{Ajami:2003aa,Ajami:2006aa,Schaller:2014aa}  
and expanded porphyrins.\cite{Stepien:2007aa,Tanaka:2008aa,Park:2008aa,Sankar:2008aa,Tokuji:2009aa}

Further advancement in the making of M{\"o}bius aromatic compounds was marked 
by the recent breakthrough in the synthesis of highly strained carbon nanobelts (CNBs). \cite{Cheung:2020aa,Guo:2021aa,Li:2021aa,Imoto:2023aa,Zhang:2023aa} 
In 2022, Segawa and coworkers \cite{Segawa:2022aa} synthesized and isolated the first M{\"o}bius CNB  
using a rational synthetic route.
The molecule is topologically characterized by a twist in an armchair-type CNB (25,25) \cite{Wang:2024aa} 
containing 50 fully fused benzenoid rings. 
One year later, Fan et al. \cite{Fan:2023aa} reported the synthesis of a triply twisted M{\"o}bius CNB 
consisting of 24 cata-condensed\cite{Ehrenhauser:2015aa} benzenoid rings. 
Compared to prior monocyclic M{\"o}bius aromatic compounds with a single-stranded $\pi$-conjugation framework, 
these two synthesized CNBs present the first examples of double-stranded conjugated systems with M{\"o}bius topology. 
Apart from the fundamental significance in M{\"o}bius aromaticity, 
they exhibit interesting optical and chiroptical properties. 
Both molecules emit fluorescence due to their fully conjugated electronic structure.\cite{Segawa:2022aa,Fan:2023aa}
The twisted M{\"o}bius structure imparts topological chirality, 
and notably, the triply twisted [24]M{\"o}bius CNB  
displays a large absorption dissymmetry factor and remarkable circularly polarized luminescence brightness, 
suggesting its potential application in chiroptics.\cite{Fan:2023aa}

In light of these recent advancements in the atomically precise synthesis of M{\"o}bius CNBs, 
we find it both intriguing and essential to conduct a theoretical survey on the $\pi$ electronic structures of these double-stranded M{\"o}bius conjugation systems. 
Such an investigation could not only deepen our understanding of their stability and physicochemical properties, 
but also help extend the Heilbronner aromaticity rules from monocyclic to polycyclic M{\"o}bius molecules. 
To start with, we shall employ the simple H{\"u}ckel molecular orbital (SHMO) theory \cite{Huckel70,Huckel72,Huckel76,Huckel83} 
(or the tight-binding method in physics terminology). 
As one of the simplest quantum models for $\pi$-conjugated systems, 
the SHMO theory relies solely on atomic connectivity,  
making it both computationally efficient and more physically interpretable
compared to more sophisticated quantum chemical approaches, such as semiempirical and DFT methods. 
Previous studies have demonstrated that 
the SHMO theory effectively models various nonplanar $\pi$-conjugated carbon systems, 
offering an in-depth understanding of the stability of charged fullerenes,\cite{Wang10,PhysRevA.80.033201,wang2015,Wang:2018uv} 
infinitenes,\cite{Du:2023aa} generalized kekulenes,\cite{kek_clr} clarenes,\cite{kek_clr} and CNBs,\cite{Wang:2024aa} 
as well as the regioselectivity in exohedral additions for neutral fullerenes\cite{Wang:2017xy,Wang:2018uv} and carboncones.\cite{Chen:2022aa} 

Regarding the specific case of M{\"o}bius CNBs, 
the applicability of the SHMO theory to these double-stranded, half-twisted conjugated systems remains unclear.
In a M{\"o}bius nanobelt the overlap between $\pi$ AOs would not only be considerably reduced,\cite{Rzepa:2005aa,Herges:2006aa}
but would also vary significantly across the molecule 
due to the uneven distribution of structural deformation caused by the single half-twist in the M{\"o}bius belt, 
which may challenge the simple approximations made in the SHMO model. 
Furthermore, addressing the phase inversion in AO basis is a nontrivial issue 
for double-stranded conjugated systems. 
For monocyclic polyenes with M{\"o}bius conjugation, 
the phase inversion can be arbitrarily placed at any position in the monocycle, 
as all C--C bonds are {\em topologically} equivalent. 
In a M{\"o}bius CNB, however, topologically nonequivalent C--C bonds exist, 
raising an interesting and fundamental question: 
do we obtain different or identical results from the SHMO theory 
when placing the phase inversion at topologically distinct bonds?
The answer is not as obvious as it might appear. 
Explicit SHMO solutions were reported long ago \cite{Gutman:1996aa,Fowler:2002aa},
but only for the simplest case of M{\"o}bius CNBs,
which consist of linear polyacenes where all benzenoid rings are topologically equivalent. 
However, to the best of our knowledge, 
the vast majority of M{\"o}bius CNBs with diverse ring annulation patterns, 
including those that have been experimentally synthesized, remain unexplored,
leaving the general principles governing their electronic structures yet to be established.

In this paper, we mathematically demonstrate that 
properly varying the sign inversion positions in any M{\"o}bius CNB can result in    
signature similarity transformations of the adjacency matrix. 
As a result, the phase inversion in AOs can be assigned to different pairs of C--C bonds 
in a quite arbitrary manner, 
yielding the same H{\"u}ckel $\pi$ electronic structure for M{\"o}bius CNBs. 
We then apply the SHMO theory to representative M{\"o}bius CNBs 
with different macrocyclization types and various molecular sizes, 
including the experimental product [50]M{\"o}bius CNB.\cite{Segawa:2022aa} 
Our results show that the SHMO theory reliably produces $\pi$ molecular orbitals (MOs), 
$\pi$ bond orders, and HOMO--LUMO gaps in striking agreement with those obtained 
from more elaborate DFT or semiempirical DFT calculations. 
Additionally, we propose a method for constructing MOs with ideally pure $\pi$ character 
in nonplanar conjugated systems, 
thereby facilitating the comparison of MOs between DFT and SHMO methods.

The remainder of the paper is organized as follows.  
In Section \ref{sec:huckel}, we introduce the SHMO theory for M{\"o}bius CNBs 
and prove the sign inversion theorem for adjacency matrix. 
In Section \ref{sec:support}, we validate the general applicability of the SHMO theory to M{\"o}bius CNBs, 
by comparing its results with those from more accurate quantum chemical calculations. 
The subsequent section concludes the study with perspectives for future work.  
We describe the computational methods in Section \ref{sec:methods}. 
We provide additional mathematical and technical details in the Appendices.

\section{\label{sec:huckel}
SHMO theory for M{\"o}bius CNBs
}

\begin{figure}[h]
\includegraphics[width=0.8\textwidth]{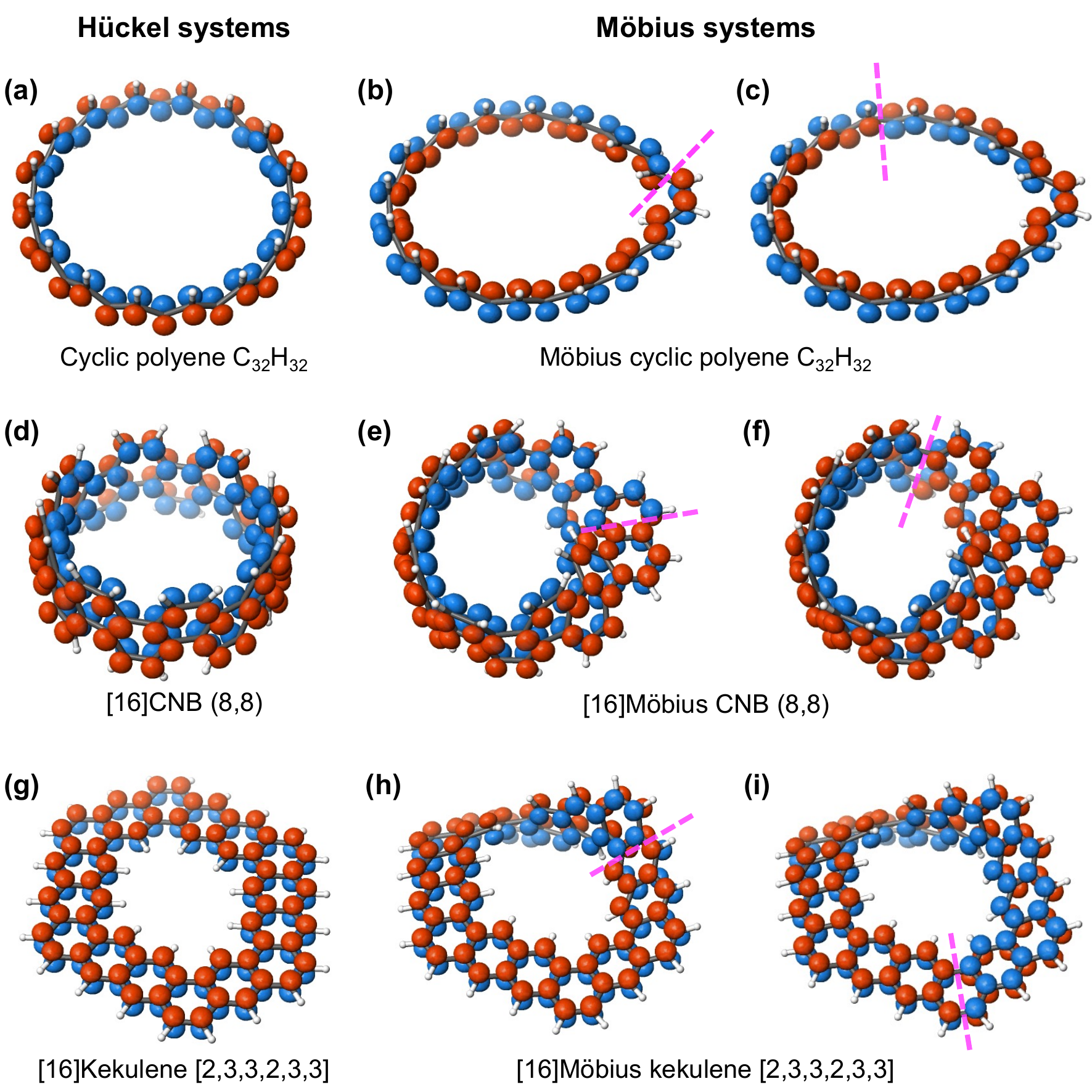}
\caption{\label{fig:orb} 
Schematic depiction of the $\pi$ AOs of carbon atoms as basis set for $\pi$-conjugated systems.
(a) Cyclic polyene C$_{32}$H$_{32}$ and (b,c) its twisted M{\"o}bius structure, 
(d) [16]CNB (8,8), (e,f) [16]M{\"o}bius CNB (8,8), 
(g) [16]kekulene [2,3,3,2,3,3] and (h,i) its M{\"o}bius counterpart.
The same M{\"o}bius $\pi$ system can have the sign inversion in $\pi$ AOs located at different positions 
(indicated by magenta dashed lines), 
as shown in (b) vs. (c), (e) vs. (f), and (h) vs. (i).
}
\end{figure}

For monocyclic M{\"o}bius $\pi$ systems, 
a single sign inversion occurs as we traverse around the cyclic array of AO basis 
due to the 180$^\circ$ half-twist that creates the M{\"o}bius topology (Fig.~\ref{fig:orb}b,c) 
from an untwisted, H{\"u}ckel topology (Fig.~\ref{fig:orb}a). \cite{Zimmerman:1971aa}
We indicate the phase inversion in $\pi$ AOs with a magenta dashed line in Fig.~\ref{fig:orb}b, 
taking the M{\"o}bius cyclic polyene C$_{32}$H$_{32}$ as an example. 
Consequently, within the SHMO model, 
the adjacency matrix for a monocyclic M{\"o}bius polyene remains identical to that of its untwisted, H{\"u}ckel counterpart, 
except that the matrix elements associated with the C--C bond where the phase inversion occurs 
change from 1 to $-1$. 
This is because the phase shift in AO basis inverts the sign of the corresponding Hamiltonian integral 
(resonance integral), changing it from $\beta$ to $-\beta$. 
In other words, we can represent the H{\"u}ckel Hamiltonian of a M{\"o}bius conjugated molecular system 
using the adjacency matrix of a signed graph, \cite{Harary53,Zaslavsky:1982aa,Lee:1994aa}
where each connection (edge) between atoms (nodes) is assigned a value of $-1$ 
if the atoms have opposite AO phases or a value of +1 if their AOs are in phase.
In a more realistic treatment,\cite{Heilbronner:1964aa,Rzepa:2005aa,Herges:2006aa} 
the resonance integral, $\beta$, is further modulated by a factor of $\cos(\pi/N)$ ($N$ being the number of atoms in the monocyclic $\pi$ system) 
to account for the reduced overlap between $\pi$ AOs due to the twisted carbon framework. 
Nevertheless, this modification does not alter 
the direct relationship between the H{\"u}ckel Hamiltonian matrix and the adjacency matrix, 
since all resonance integrals are universally scaled.
Moreover, a notable observation is that 
because of the cyclic symmetry in {\em topology} ({\em not geometry}), 
the adjacency matrix is isomorphism invariant   
when changing the position of sign inversion from one C--C bond to another 
(e.g., comparing Fig.~\ref{fig:orb}b and Fig.~\ref{fig:orb}c). 

In this work, we focus on double-stranded M{\"o}bius CNBs,\cite{Wang:2024aa} 
structurally defined as closed hydrocarbon macrocycles of cata-condensed\cite{Ehrenhauser:2015aa} benzenoid rings, 
where each hexagonal ring is fused to two others, with no carbon atoms shared among three rings, 
as exemplified by the structures shown in Fig.~\ref{fig:orb}d--i. 
Due to their double-stranded carbon framework, 
at least two C--C bonds must be broken to open the molecular macrocycle.\cite{Segawa:2016aa,Imoto:2023aa} 
There are two macrocyclization types of CNBs, the radial type and the parallel type, 
as exemplified by [16]CNB (8,8)\cite{Wang:2024aa} (Fig.~\ref{fig:orb}d) 
and [16]kekulene [2,3,3,2,3,3]\cite{kek_clr} (Fig.~\ref{fig:orb}g), respectively. 
The readers can refer to refs. \citenum{Du:2023aa,kek_clr} 
for an explanation of the nomenclature of extended kekulenes. 
In the radial type,  
the normals of the benzenoid planes (and thus the $\pi$ AOs) are oriented radially 
with respect to the molecular macrocycle. 
In contrast, in the parallel type, the benzenoid planes are aligned more or less parallel with the macrocyclic plane. 
Radial and parallel types of M{\"o}bius CNBs can be obtained by 
introducing a half-twist into the respective CNBs. 
Fig.~\ref{fig:orb}e,f and Fig.~\ref{fig:orb}h,i show the radial-type [16]M{\"o}bius CNB (8,8) 
and the parallel-type [16]M{\"o}bius kekulene [2,3,3,2,3,3], respectively. 

Similar to monocyclic polyenes, 
we can likewise apply the SHMO theory to M{\"o}bius CNBs 
by introducing sign inversion in the adjacency matrix. 
However, this raises the question: 
Does the adjacency matrix preserve its eigenvalues and eigenvectors with respect to variations in the position of the sign inversion?
For instance, do the adjacency matrices associated with the two different AO phase arrangements  
shown in Fig.~\ref{fig:orb}e and Fig.~\ref{fig:orb}f have identical spectra and yield the same set of H{\"u}ckel molecular orbitals? 
The answer to this question is not straightforward. 
Unlike monocyclic polyenes, 
the positions of phase inversion in M{\"o}bius CNBs are not necessarily topologically equivalent. 
This is evident from the differences between Fig.~\ref{fig:orb}e and Fig.~\ref{fig:orb}f,
as well as between Fig.~\ref{fig:orb}h and Fig.~\ref{fig:orb}i, 
where the phase inversion positions are marked by magenta dashed lines. 

In the following, we show that, at least for double-stranded M{\"o}bius CNBs, 
the adjacency matrix satisfies a specific sign inversion theorem. 
Prior to presenting the theorem, 
we need to define the concepts of half-cuts and minimum half-cuts in the context of molecular graphs of CNBs. 
A half-cut of a CNB graph is a way of removing a set of edges 
such that the macrocyclic structure is opened but the resultant graph remains connected.
In comparison, a cut of a graph is a partition of vertices into two disjoint subsets. 
A minimum half-cut is then defined as the half-cut that removes the fewest edges. 
For a double-stranded M{\"o}bius CNB a minimum half-cut involves removing two edges, 
as opening the CNB macrocycle necessitates the cleavage of at least two C--C bonds. 

\textbf{Theorem:} 
\textit{For the graph of any double-stranded M{\"o}bius CNB, 
varying the sign inversion positions corresponding to any minimum half-cut 
preserves the eigenvalues of the adjacency matrix and may flip only the signs of its eigenvectors.}

\textit{Proof:}
We begin by defining the \textit{negation operation} on node $p$ in a graph 
such that it negates all elements in both row $p$ and column $p$ of the adjacency matrix, $\textbf{A}$. 
It is straightforward to observe that this negation operation is a signature similarity transformation: 
\begin{equation}\label{eq:neg}
\hat{N}_p \mathbf{A} = \mathbf{S}_p^{-1} \mathbf{A} \mathbf{S}_p = \mathbf{S}_p \mathbf{A} \mathbf{S}_p
\end{equation}
where $\hat{N}_p$ is the negation operator, and $\mathbf{S}_p$ is a signature matrix, 
a diagonal matrix with the $p$th diagonal entry as $-1$ and all other diagonal entries as 1 
(see Fig.~\ref{fig:sign_inv}b).
A detailed derivation can be found in Appendix \ref{app:sigmat}. 

\begin{figure}
\includegraphics[width=0.5\textwidth]{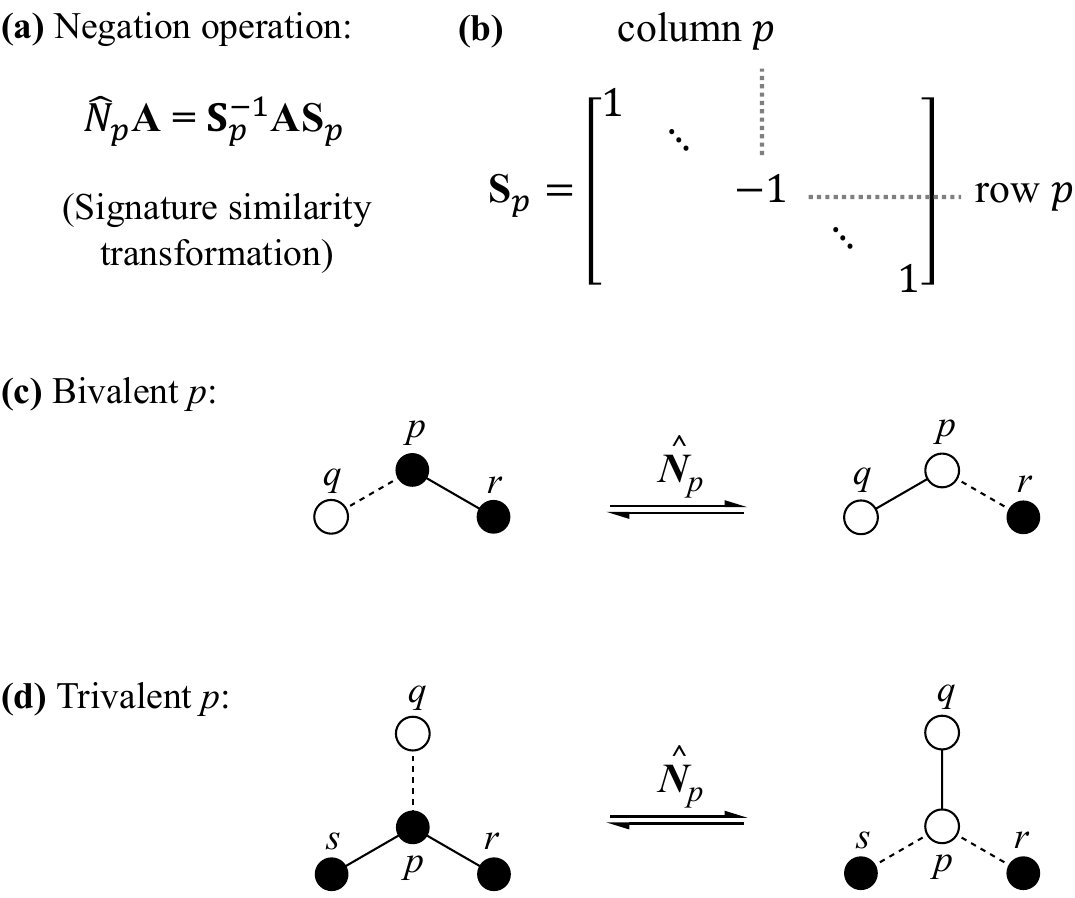}
\caption{\label{fig:sign_inv} 
(a) A negation operation on node (atom) $p$ in a molecular graph, $\hat{N}_p$, performs a signature similarity transformation 
on the adjacency matrix $\mathbf{A}$. 
(b) The signature matrix, $\mathbf{S}_p$, associated with the negation operator, $\hat{N}_p$. 
(c) Reversible negation operation on a bivalent node $p$ inverts the phase of its AO 
and hence the position where the phase shift takes place. 
(d) Negation operation on a trivalent node $p$. 
AOs are represented by white and black circles. 
The C--C bonds connecting two in-phase atoms are depicted using solid lines, 
while the two opposite-phase atoms are connected by a dashed line. 
}
\end{figure}

In CNB graphs, 
nodes are either bivalent, representing carbon atoms bonded to two other carbon atoms and one hydrogen atom, 
or trivalent, corresponding to carbon atoms bonded to three other carbon atoms. 
Fig.~\ref{fig:sign_inv}c illustrates a bivalent node $p$ having an edge ($p\text{---}q$) with sign inversion 
(indicated by a dashed line connecting black and white circles, which denote opposite AO phases). 
When we perform a negation operation on $p$, 
the sign inversion migrates from edge $p\text{---}q$ to edge $p\text{---}r$. 
Conversely, reapplying the same negation operation restores the resulting graph to the original graph 
(right to left in Fig.~\ref{fig:sign_inv}c). 
A similar reversible negation can be applied to a trivalent node with a sign inversion edge, 
as shown in Fig.~\ref{fig:sign_inv}d. 
In this case, the number of sign inversion edges may either increase or decrease. 

\begin{figure}
\includegraphics[width=0.65\textwidth]{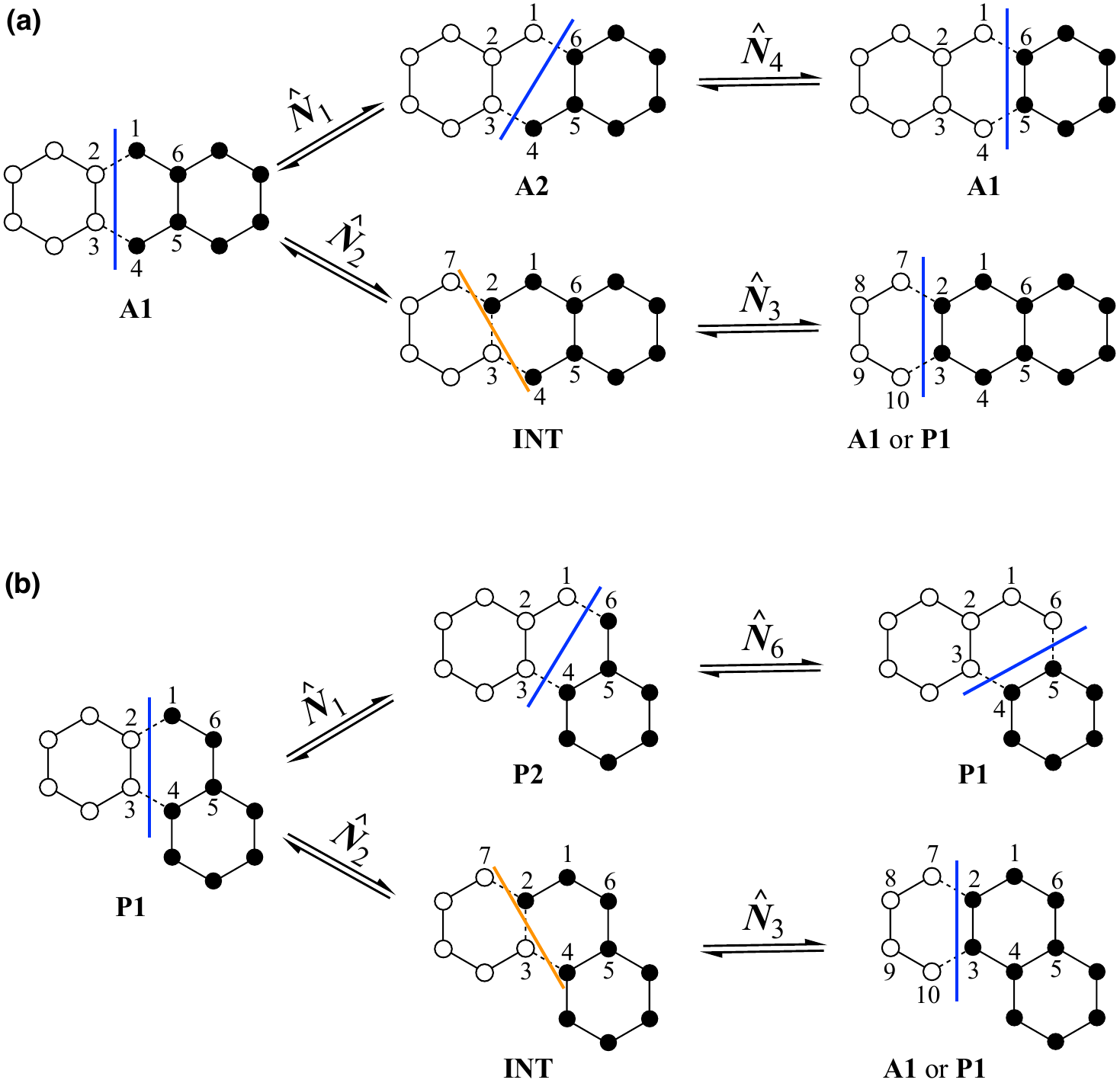}
\caption{\label{fig:theorem} 
(a) Interconversion between minimum half-cuts, \textbf{A1}, \textbf{A2}, and \textbf{P1}, by
direct or successive negation operations on appropriate atoms. 
(b) Interconversion between minimum half-cuts, \textbf{P1}, \textbf{P2}, and \textbf{A1}. 
In the minimum half-cuts, the positions of orbital phase shift are indicated by dashed lines, 
with a blue line drawn through. 
The graphs labeled as `INT' does not correspond to a minimum half-cut, 
since they have three bonds with a phase shift, which is indicated by an orange line.
}
\end{figure}

For any double-stranded CNBs, as demonstrated in Appendix \ref{app:cuts}, 
there are only four possible minimum half-cut patterns, namely, 
\textbf{A1}, \textbf{A2}, \textbf{P1}, and \textbf{P2} (see Fig.~\ref{fig:theorem} or Fig.~\ref{fig:patterns_cut}). 
We then show that starting from any of the four minimum half-cut patterns, 
it is possible to reach all the others by directly or successively applying negation operations 
on the appropriate nodes. 
To be more specific, in Fig.~\ref{fig:theorem}a, we begin with a minimum half-cut pattern \textbf{A1}, 
which has two sign inversion edges, 1---2 and 3---4 (indicated by a blue line in Fig.~\ref{fig:theorem}a). 
A negation operation on node 1, $\hat{N}_1$ (cf. Fig.~\ref{fig:sign_inv}c for negation on a bivalent node), 
directly leads to a minimum half-cut pattern, \textbf{A2}, 
where the sign inversion edge has moved from 1---2 to 1---6. 
Continuing a negation operation on node 4, $\hat{N}_4$, 
we obtain a third graph with a different minimum half-cut, which still belongs to pattern \textbf{A1}. 
Alternatively, we can apply the negation operation on node 2 in the starting graph \textbf{A1}, 
as shown in the lower route in Fig.~\ref{fig:theorem}a. 
The resultant graph (labeled as `INT' in Fig.~\ref{fig:theorem}a) contains three sign inversion edges 
(indicated by an orange line) and thus does not correspond to a minimum half-cut. 
Serving as an intermediate graph, 
it can be further transformed into a minimum half-cut pattern through a negation on node 3. 
As illustrated in Fig.~\ref{fig:theorem}a, 
the resulting minimum half-cut pattern can be regarded as either \textbf{A1} or \textbf{P1}, 
depending on where the leftmost benzenoid ring (2---7---8---9---10---3) 
annulates with a neighboring ring from left side 
(sharing the edge of either 7---8, or 8---9, or 9---10). 

Fig.~\ref{fig:theorem}b shows that a minimum half-cut pattern \textbf{P1}, 
can be reversibly transformed into \textbf{P2} (via $\hat{N}_1$) and subsequently to \textbf{P1} (via $\hat{N}_6$), 
Alternatively, \textbf{P1} can be transformed into \textbf{A1} or \textbf{P1} 
through successive steps via $\hat{N}_2$ and $\hat{N}_3$ (see the lower pathway in Fig.~\ref{fig:theorem}b).
To sum up, we have demonstrated that 
all four possible minimum half-cut patterns for double-stranded CNBs can be interconverted  
through direct or successive negation operations, each being a signature similarity transformation. 
As a result, 
the adjacency matrices for minimum half-cut patterns share the same set of eigenvalues,  
with their associated eigenvectors being either identical or flipped (in opposite directions). 
Thus, we have proven the theorem.

This theorem has fundamental chemical significance: 
the AO phase inversion can be placed at any position around the nanobelt, 
provided that the cleavage of two C---C bonds within the same benzenoid ring open the M{\"o}bius macrocycle. 
The corresponding simple H{\"u}ckel wave function (in the form of a Slater determinant), 
and hence the $\pi$ electronic structure, remains invariant regardless of the choice of sign inversion positions, 
even if those positions are geometrically or topologically nonequivalent. 
This invariance arises from the fact that 
the signature similarity between adjacency matrices guarantees isospectrality (though the converse is not necessarily true), 
thereby preserving the H{\"u}ckel MO energies. 
The only change in eigenvectors under signature similarity transformations is a potential flip in their directions, 
leading to a sign change of the corresponding H{\"u}ckel MO wave functions. 
Therefore, there is no need to concern ourselves with selecting the most ``appropriate'' position 
for phase inversion along the M{\"o}bius nanobelt, 
even though intuition might suggest that the site of the half-twist would naturally be the ``best'' choice. 

Lastly, it is necessary to note that the definition of Coulson bond order \cite{Coulson:1997aa,Levine_QC} 
within the SHMO theory should be adapted to M{\"o}bius conjugated systems as follows: 
\begin{equation}\label{eq:bo}
p_{rs} = \sum_{i=1}^{N} n_i s_{rs} c_{ri} c_{si}
\end{equation} 
where $p_{rs}$ is the bond order between two adjacent carbon atoms, $r$ and $s$; 
the sum runs over all $N$ H{\"u}ckel MOs; $n_i$ is the occupancy of the $i$th MO; 
$c_{ri}$ and $c_{si}$ are the coefficients for the $i$th MO associated with atoms $r$ and $s$, respectively;
and the sign factor $s_{rs}$ takes a value of $-1$ if there is a phase inversion between AOs of $r$ and $s$, 
and $s_{rs} = 1$ otherwise.

\section{\label{sec:support}
Corroboration by sophisticated quantum chemical calculations
}

\begin{figure}[h]
\includegraphics[width=0.6\textwidth]{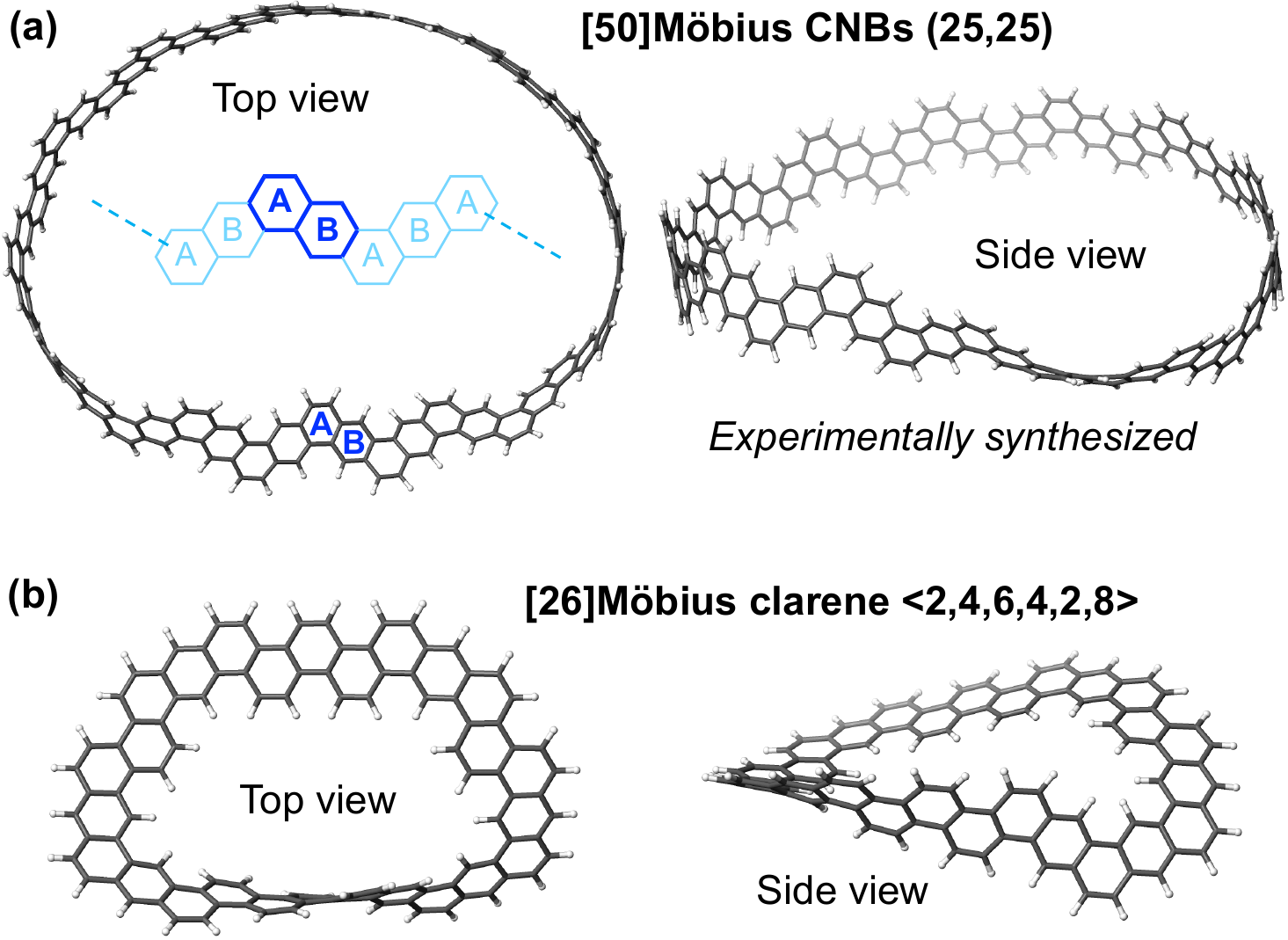}
\caption{\label{fig:str} 
Top and side views of DFT optimized structures of 
(a) the experimentally synthesized [50]M{\"o}bius CNB (25,25) \cite{Segawa:2022aa} and 
(b) the hypothetical [26]M{\"o}bius clarene <2,4,6,4,2,8>. 
In (a), the topologically repeating unit of rings is highlighted in blue. 
The two topologically nonequivalent rings are labels as A and B. 
}
\end{figure}

To verify the adequacy and efficacy of the SHMO theory 
in describing the $\pi$ electronic structure of M{\"o}bius CNBs, 
we compare the SHMO theory predictions of orbital shapes, $\pi$ bond orders, and HOMO--LUMO gaps, 
with the results obtained from DFT and GFN2-xTB (xTB for short hereafter) \cite{Grimme:2017wn,Bannwarth:2019th} calculations. 
We consider two representative M{\"o}bius CNBs for testing: 
the radial-type [50]M{\"o}bius CNB (25,25) and the parallel-type [26]M{\"o}bius clarene <2,4,6,4,2,8> 
(see refs. \citenum{Du:2023aa,kek_clr} for the nomenclature of clarenes), 
as depicted in Fig.~\ref{fig:str}a and Fig.~\ref{fig:str}b, respectively. 
Recently synthesized in the lab \cite{Segawa:2022aa}, 
the first example consists of 50 benzenoid rings in an armchair arrangement. 
As shown in Fig.~\ref{fig:str}a, 
there are only two topologically nonequivalent rings (labeled as A and B). 
The other test example is a hypothetically designed M{\"o}bius nanobelt, 
structurally obtained by half-twisting a [26]clarene <2,4,6,4,2,8> molecule.\cite{kek_clr}
This molecule, as illustrated in Fig.~\ref{fig:str}b, is topologically less symmetric 
and exhibits geometric deformation due to its smaller size.

\subsection{\label{sec:piCMO}
Construction of DFT $\pi$-CMOs in M{\"o}bius systems
}

Due to the nonplanar geometry with uneven distribution of the twist in a M{\"o}bius macrocyclic molecule, 
the frontier canonical MOs (CMOs) obtained from DFT calculations 
often have a mix of $\pi$ and $\sigma$ characters. 
As pointed out by Rzepa \cite{Rzepa:2005aa} and Herges \cite{Herges:2006aa}, 
in several M{\"o}bius cyclacenes the HOMO and LUMO have little distribution 
in the region around the geometric twist. 
Lower-lying $\pi$-character CMOs exhibit a considerable contribution from $\sigma$ bonding 
(see Fig. 25 in ref. \citenum{Herges:2006aa}). 
Similarly, in the case of M{\"o}bius CNBs, $\sigma$--$\pi$ mixing is also observed, 
as demonstrated by the HOMO$-95$ of [26]M{\"o}bius clarene <2,4,6,4,2,8>, shown in Fig.~\ref{fig:mos_2}. 

Therefore, the original CMOs obtained from a DFT computation may not be ideal  
for direct comparison with the pure $\pi$ MOs derived from SHMO calculation. 
An effective solution is to separate the $\pi$ electrons from the $\sigma$ and inner-shell electrons in a given molecule. 
The idea is to first transform the CMOs into localized MOs (LMOs) through unitary rotations, 
ensuring that the single Slater determinant for the entire wave function remains invariant. 
Next, we isolate all the $\pi$-LMOs that define the $\pi$ subsystem  
and then recombine them, via unitary transformation, to construct the CMOs for the $\pi$ subsystem 
(referred to as $\pi$-CMOs hereafter).

In the following discussion, we assume a closed-shell electronic structure for the systems in question. 
We adopt the Pipek--Mezey localization \cite{doi:10.1063/1.456588} in the natural atomic orbital (NAO) \cite{nao} basis, 
as this scheme provides a clear $\sigma$--$\pi$ separation \cite{doi:10.1063/1.456588} 
compared to other schemes such as Foster--Boys \cite{RevModPhys.32.296,RevModPhys.32.300} and
Edmiston--Ruedenberg \cite{RevModPhys.35.457,doi:10.1063/1.1701520} localizations. 
The obtained coefficient matrix of LMOs, expressed in the NAO basis, 
is denoted as $\mathbf{C}^\mathrm{LMO}_\mathrm{NAO}$. 
We then separate all $N/2$ (where $N$ is the number of carbon atoms) occupied $\pi$ LMOs, 
denoted as $\Omega_i \; (i = 1,2,\dots,N/2)$, 
from the $\sigma$ and inner-shell LMOs. 
The occupied $\pi$ LMOs usually correspond to the highest occupied LMOs,
which can be further verified by visual inspection of orbital shape and distribution. 
The Fock matrix in the complete LMO basis is given by 
\begin{equation}\label{eq:fockmat}
\mathbf{F}_\mathrm{LMO} = \mathbf{C}^\mathrm{LMO \dagger}_\mathrm{NAO} \mathbf{F}_\mathrm{NAO} \mathbf{C}^\mathrm{LMO}_\mathrm{NAO}
\end{equation}
where $\mathbf{F}_\mathrm{NAO}$ is the Fock matrix in the NAO basis 
and $\dagger$ denotes conjugate transpose. 
By extracting from $\mathbf{F}_\mathrm{LMO}$ the subblock associated with the occupied $\pi$ LMOs, 
we obtain the Fock matrix for the $\pi$ subsystem, denoted as $\mathbf{F}_\mathrm{LMO}^{\pi}$. 
Diagonalization of $\mathbf{F}_\mathrm{LMO}^{\pi}$ yields a set of eigenvectors and eigenvalues. 
While the eigenvalues, $\{E_i^{\pi}\; (i = 1, 2, \dots, N/2)\}$, 
can be regarded as the energies of the occupied $\pi$-CMOs, 
the eigenvectors represent the coefficients of the occupied $\pi$-CMOs in the $\pi$ LMO basis, 
denoted as $\mathbf{C}^{\pi}_\mathrm{LMO}$. 
To obtain the $\pi$-CMO coefficients in the NAO basis, we perform the following basis transformation: 
\begin{equation}
\mathbf{C}^{\pi}_\mathrm{NAO} = \mathbf{C}^{\pi\text{-LMO}}_\mathrm{NAO} \mathbf{C}^{\pi}_\mathrm{LMO}
\end{equation}
where $\mathbf{C}^{\pi\text{-LMO}}_\mathrm{NAO}$ is the coefficient matrix of the occupied $\pi$-CMOs 
in the NAO basis, as extracted from $\mathbf{C}^\mathrm{LMO}_\mathrm{NAO}$. 
Eventually, one may wish to express the occupied $\pi$-CMOs in the original AO basis as $\mathbf{C}^{\pi}_\mathrm{AO}$, 
which can be exported to a \textsc{Gaussian} formchk file for visualization of these $\pi$ orbitals:
\begin{equation}
\mathbf{C}^{\pi}_\mathrm{AO} = \mathbf{C}^\mathrm{NAO}_\mathrm{AO} \mathbf{C}^{\pi}_\mathrm{NAO}
\end{equation}
where $\mathbf{C}^\mathrm{NAO}_\mathrm{AO}$ is the transformation matrix 
from the original AO basis to the NAO basis set.\cite{nao} 

To compute the Wiberg $\pi$ bond indices \cite{WIBERG19681083}, 
we first build the first-order reduced density matrix in the NAO basis for the $\pi$ system 
from the coefficients of all occupied $\pi$ LMO: 
\begin{equation}
\mathbf{D}^{\pi}_\mathrm{NAO} = \mathbf{C}^{\pi\text{-LMO}}_\mathrm{NAO} \mathbf{C}^{\pi\text{-LMO} \dagger}_\mathrm{NAO}
\end{equation}
The Wiberg $\pi$ bond index between any two atoms, $r$ and $s$, is then calculated as 
\begin{equation}
b_{rs} = \mathrm{Tr}( \mathbf{D}^{\pi}_{\mathrm{NAO},rs} \mathbf{D}^{\pi}_{\mathrm{NAO},sr} )
\end{equation}
where $\mathrm{Tr}(\;)$ denotes the matrix trace; 
$\mathbf{D}^{\pi}_{\mathrm{NAO},rs}$ and $\mathbf{D}^{\pi}_{\mathrm{NAO},sr}$ 
are the interatomic blocks of the density matrix $\mathbf{D}^{\pi}_\mathrm{NAO}$ 
between atoms $r$ and $s$. \cite{Wang:2018aa}

\subsection{\label{sec:comp_piCMO}
Comparison of HMOs and DFT $\pi$-CMOs 
}

\begin{figure}[h]
\includegraphics[width=\textwidth]{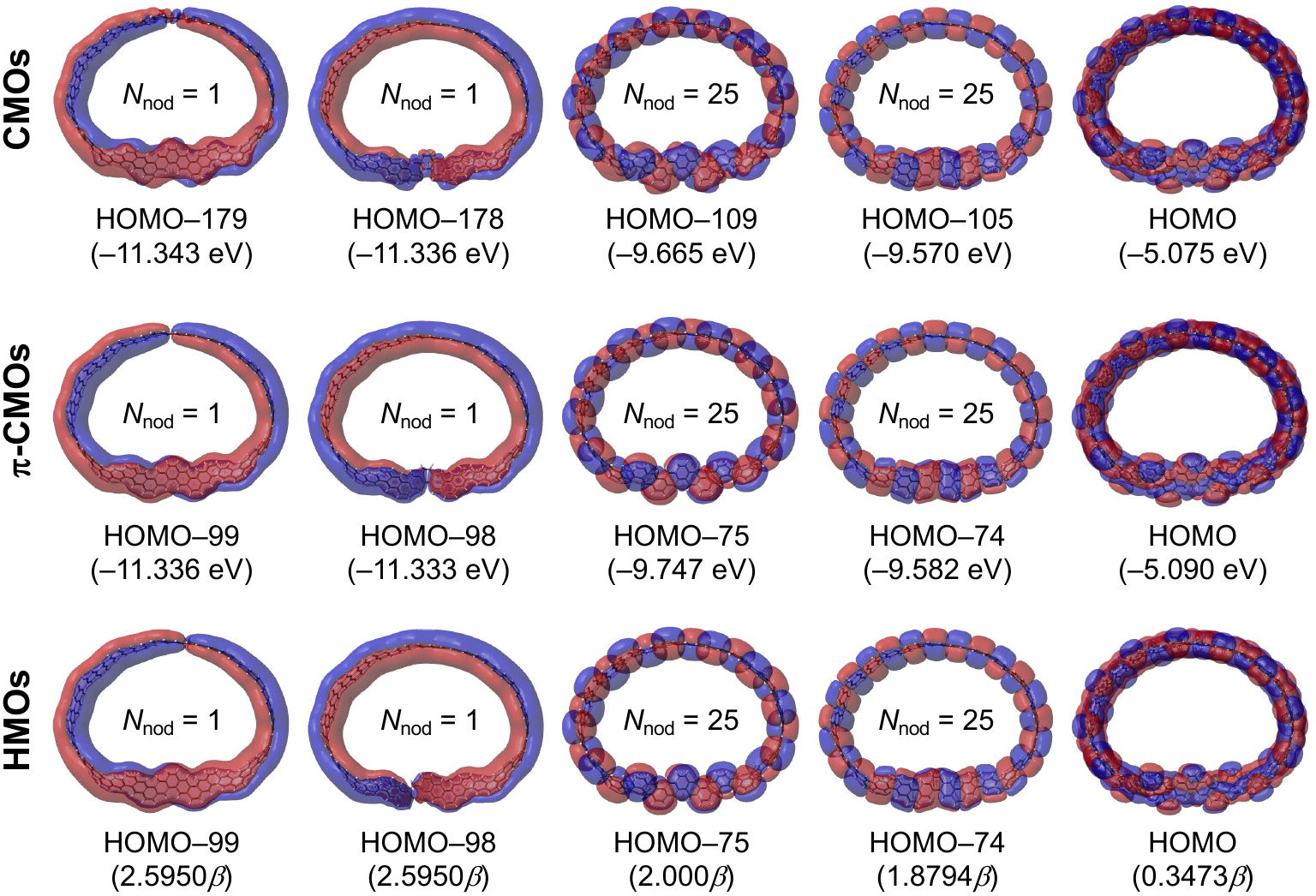}
\caption{\label{fig:mos_1} 
Comparison of DFT-calculated CMOs and $\pi$-CMOs with HMOs 
for the experimentally synthesized [50]M{\"o}bius CNB (25,25). \cite{Segawa:2022aa}
The lowest $\pi$ MOs, HOMO, and a pair of representative $\pi$ MOs are selected. 
Orbital isosurfaces are plotted with an isovalue of 0.0002. 
Orbital energies are given in parentheses. 
The number of nodal planes in the $\pi$-conjugation direction ($N_\mathrm{nod}$) is indicated for the lower MOs. 
}
\end{figure}

Fig.~\ref{fig:mos_1} compares the DFT-computed CMOs and $\pi$-CMOs with the HMOs 
for the synthesized [50]M{\"o}bius CNB (25,25). \cite{Segawa:2022aa} 
To visualize the isosurfaces of HMOs in 3D space, 
we constructed the corresponding wave functions 
as linear combinations of idealized carbon's 2p$_z$ AOs, using the coefficients obtained from SHMO calculations 
(see Appendix \ref{app:MOs} for details). 
As shown, the plotted HMOs agree remarkablely well with the DFT CMOs and $\pi$-CMOs 
in terms of both orbital shapes and wave function distribution over the molecule. 
While one can still discern the slight difference between the CMOs and the corresponding HMOs 
(especially for the two lowest-lying MOs), 
the $\pi$-CMOs and the HMOs bear a striking resemblance to each other. 

The lowest two HMOs, HOMO$-99$ and HOMO$-98$, are doubly degenerate. 
In a M{\"o}bius $\pi$ system, 
the inherent phase inversion in AO basis means that 
the lowest (most bonding) $\pi$ MOs must feature a nodal plane along the $\pi$-conjugation direction. 
This contrasts with a H{\"u}ckel $\pi$ system, 
where the lowest MO acquires an all-bonding combination of $\pi$ AOs, 
as seen in benzene, one of the simplest examples.
In the particular case of [50]M{\"o}bius CNB (25,25), 
because of its high topological symmetry 
(i.e., possessing 25 topologically equivalent units of  rings, as shown in Fig.~\ref{fig:str}), 
the location of the unavoidable nodal plane is quite arbitrary, with as many as 25 different possible positions.  
To eliminate such arbitrariness in the total wave function 
(as justified by the sign inversion theorem in Section \ref{sec:huckel}), 
it is essential to have two degenerate lowest-energy MOs with nodal planes that are perpendicular to each other  
(see the two HMOs in the bottom-left corner in Fig.~\ref{fig:mos_1}).  
This ensures that any recombination of these MOs will yield different pairs of two-fold MOs 
with mutually perpendicular nodal planes located at varying positions. 
Therefore, the double degeneracy of the lowest HMOs of [50]M{\"o}bius CNB (25,25) 
fundamentally arises from the non-orientable nature of the M{\"o}bius topology 
and the the molecule's {\em topological} symmetry. 
This phenomenon of double degeneracy due to topological requirements also occurs in other low-lying HMOs.  
For example, the lowest 24 HMOs (namely, from HOMO$-99$ to HOMO$-76$) are all doubly degenerate. 
However, the subsequent HMOs, HOMO$-75$ and HOMO$-74$ (see Fig.~\ref{fig:mos_1}), 
are not degenerate, with energy eigenvalues of $2.000 \beta$ and $1.8794 \beta$, respectively. 
The breakdown in degeneracy for these HMOs can be explained as follows:  
With 25 nodal planes, there are 25 alternately signed lobes in each of these two HMOs. 
Each lobe of the HMO encompasses one ring in the center and another at its two ends, 
given the total of 50 benzenoid rings in the nanobelt.
More specifically, in HOMO$-75$ ring A is centered within each lobe, 
whereas ring B is situated between two neighboring lobes, as shown in Fig.~\ref{fig:mos_1}. 
A contrary situation is seen in HOMO$-76$: 
ring B is at the center of each lobe and ring A spans across two neighboring lobes. 
Since rings A and B are topologically nonequivalent, 
HOMO$-75$ and HOMO$-76$ are not degenerate. 

As the energy of the HMO increases, the number of nodal planes also rises. 
A common feature is that the number of nodal planes is always odd, 
as a consequence of the M{\"o}bius topology of the $\pi$ system. 
However, in higher-energy HMOs, the nodal planes become less distinct, 
with irregular boundaries forming between regions of opposite sign, 
as observed in the HOMO shown in Fig.~\ref{fig:mos_1}.

Unlike the SHMO theory relying solely on molecular topology, 
DFT calculations take into account the actual 3D geometry of the molecule. 
As a result, 
the topological degeneracy discussed earlier does not strictly apply to the DFT-computed CMOs and $\pi$-CMOs. 
In the DFT results, 
orbital degeneracy arises from {\em geometric} rather than {\em topological} symmetry, 
and the geometric asymmetry inherent in the M{\"o}bius nanobelt disrupts this degeneracy.
For instance, the lowest two DFT-calculated $\pi$-CMOs are not exactly degenerate,  
but their energy difference is negligible (only 0.003 eV, see Fig.~\ref{fig:mos_1}). 
Similarly, the corresponding DFT CMOs (HOMO$-179$ and HOMO$-178$) 
are also nearly degenerate, with an energy difference of 0.007 eV. 
In comparison, a much larger energy difference of 0.095 eV (0.165 eV) 
is observed between the pair of MOs with 25 nodal planes for the DFT-calculated CMOs ($\pi$-CMOs), 
thus supporting the HMO results discussed above. 
It is noteworthy that many CMOs with predominantly $\sigma$ character are interspersed  
between those with significant $\pi$ character. 
Therefore, the lowest $\pi$-character CMO is HOMO$-179$, instead of HOMO$-99$, 
despite there being a total of 100 occupied $\pi$ MOs for this [50]M{\"o}bius CNB. 

\begin{figure}[h]
\includegraphics[width=\textwidth]{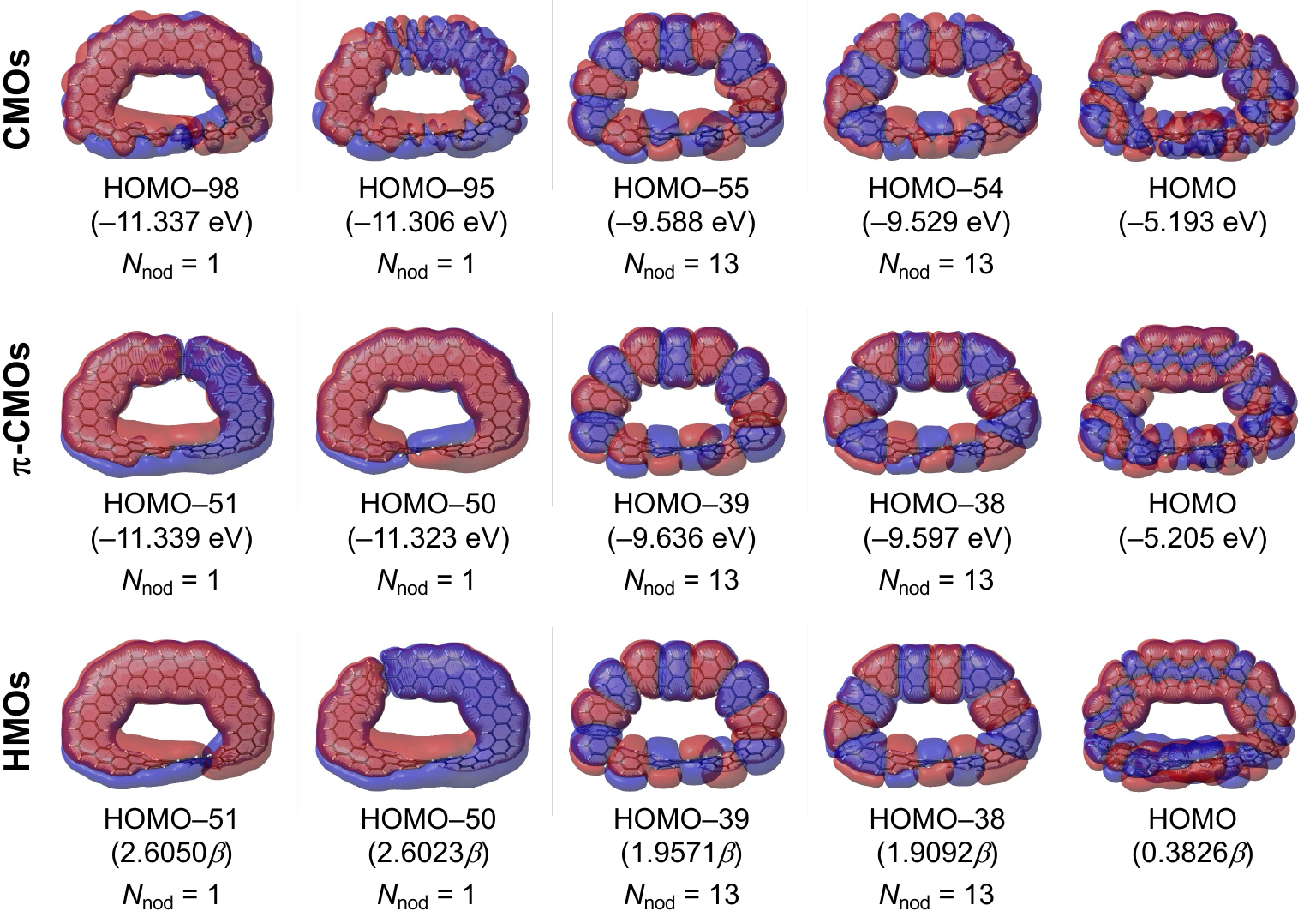}
\caption{\label{fig:mos_2} 
Idem Fig.~\ref{fig:mos_1} for [26]M{\"o}bius clarene <2,4,6,4,2,8>. 
}
\end{figure}

Now, let us analyze the $\pi$ MOs for [26]M{\"o}bius clarene <2,4,6,4,2,8>. 
As shown in Fig.~\ref{fig:mos_2}, 
there is a strong overall agreement in orbital shape and distribution 
among the HMOs, DFT-computed CMOs and $\pi$-CMOs. 
Unlike the previous case of [50]M{\"o}bius CNB (25,25), 
this [26]M{\"o}bius clarene is topologically asymmetric,  
with all rings being topologically nonequivalent. 
Consequently, all $\pi$ MOs are non-degenerate, even within the framework of SHMO theory.
Although the lowest two HMOs and the corresponding DFT $\pi$-CMO pair somehow show different ordering in energy, 
both MOs are almost degenerate, with an energy difference of only 0.0007$\beta$ and 0.016 eV, respectively. 
The lowest two DFT CMOs show the same energy ordering as the HMOs, 
but display a much large energy difference of 0.031 eV, 
which is likely due to the DFT-computed HOMO$-95$ having a considerable $\sigma$ character, 
as clearly shown in Fig.~\ref{fig:mos_2}. 
Fig.~\ref{fig:mos_2} also present a higher-energy pair of MOs with 13 nodal planes, 
and both MOs are roughly degenerate as well. 
Similar to the [50]M{\"o}bius CNB case, 
the nodal planes in the HOMO of [26]M{\"o}bius clarene are not well-defined, 
as seen in Fig.~\ref{fig:mos_2}. 
In summary, the SHMO theory successfully produces the correct appearance of $\pi$ MOs, 
in good congruence with those obtained at the more accurate DFT level.

\subsection{\label{sec:gap}
Correlation of $\pi$ bond orders and HOMO--LUMO gaps 
between HMO, DFT, and xTB methods 
}

Bond order is a crucial measure of the bonding strength between two atoms. 
In the framework of SHMO theory, 
the Coulson bond order\cite{Coulson:1997aa,Levine_QC} is conventionally used 
to describe the bonding character between two adjacent carbon atoms. 
In DFT calculations, on the other hand, 
the Wiberg bond index \cite{WIBERG19681083} is frequently used as as a measure of bond order. 
Here, we compare the Coulson bond orders from SHMO calculations 
with the DFT-computed Wiberg $\pi$ bond indices for the two M{\"o}bius CNB examples we have tested.

\begin{figure}[h]
\includegraphics[width=\textwidth]{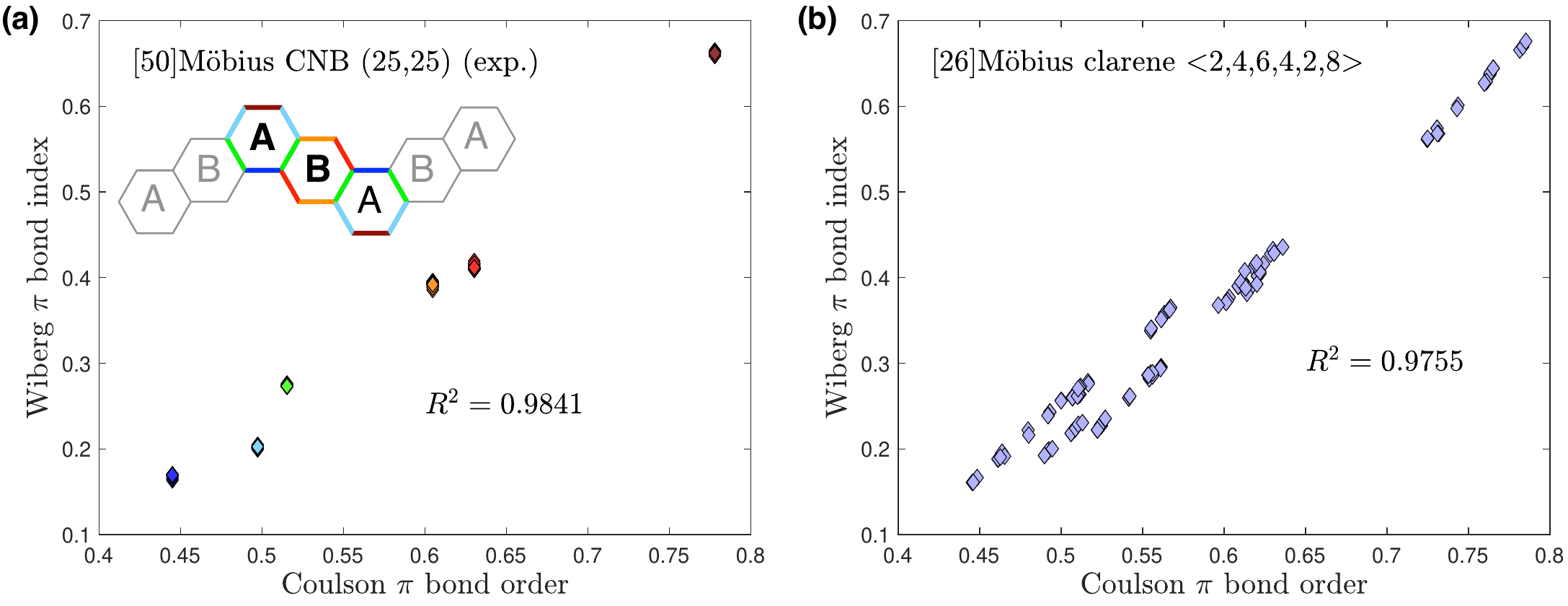}
\caption{\label{fig:bo} 
Comparison of Coulson $\pi$ bond orders from SHMO calculations with DFT-calculated Wiberg $\pi$ bond indices 
for (a) the experimentally synthesized [50]M{\"o}bius CNB (25,25) \cite{Segawa:2022aa} 
and (b) [26]M{\"o}bius clarene <2,4,6,4,2,8>. 
Squared correlation coefficient, $R^2$, is provided in each plot.
In (a), data points in different colors correspond to topologically nonequivalent C--C bonds, 
as shown in the chemical structure in the inset. 
}
\end{figure}

Fig.~\ref{fig:bo}a plots the DFT Wiberg $\pi$ bond index against the Coulson bond order 
for all C--C bonds in [50]M{\"o}bius CNB (25,25). 
There is a nice correlation between the Coulson bond order and the Wiberg $\pi$ bond index,
with a squared correlation coefficient $R^2$ of 0.9841. 
As shown in the inset, there are six topologically nonequivalent C--C bonds, 
each represented by a different color.  
Although topologically equivalent bonds are not geometrically identical in the DFT-optimized 3D structure, 
the Wiberg $\pi$ bond indices for these topologically equivalent bonds are quite similar. 
Hence, we see in Fig.~\ref{fig:bo}a that data points of the same color are clustered together. 

As for [26]M{\"o}bius clarene <2,4,6,4,2,8>, 
all 130 bonds are both topologically and geometrically distinct.
As shown in Fig.~\ref{fig:bo}b, 
the Coulson bond order correlates well with the DFT Wiberg $\pi$ bond index 
($R^2=0.9755$). 
These results further validate the effectiveness of the SHMO theory in predicting the $\pi$ bonding in M{\"o}bius CNBs.

\begin{figure}[h]
\includegraphics[width=\textwidth]{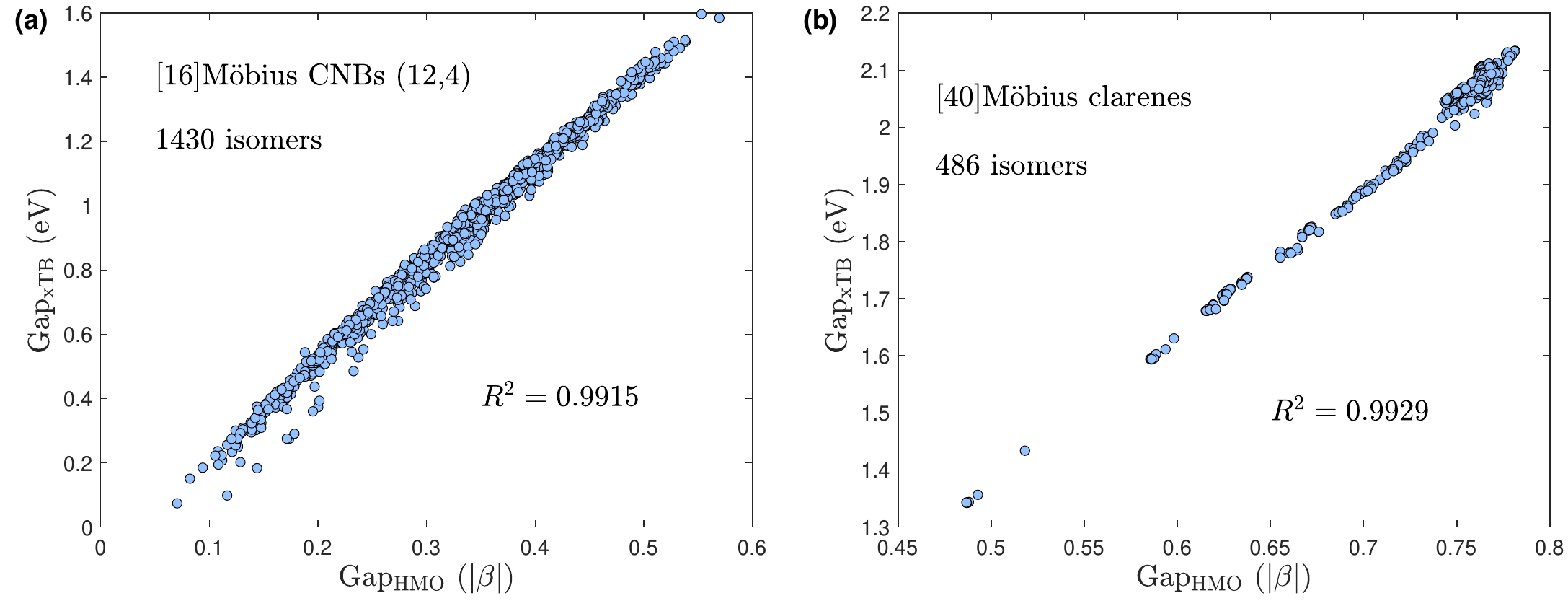}
\caption{\label{fig:gap} 
Comparison of calculated HOMO--LUMO gaps between SHMO and xTB methods 
for all generated isomers of (a) [16]M{\"o}bius CNBs (12,4) and (b) [40]M{\"o}bius clarenes. 
Squared correlation coefficient, $R^2$, is included in each plot.
}
\end{figure}

Lastly, we examine the HOMO--LUMO gaps of M{\"o}bius CNBs to assess  
the SHMO theory's quantitative predictive capability in this regard.
To obtain statistically meaningful results, 
we selected both macrocyclization types of M{\"o}bius CNBs of different molecular sizes: 
the radial-type [16]M{\"o}bius CNBs (12,4) and the parallel-type [40]M{\"o}bius clarenes. 
In both cases, 
we enumerated all isomers of the untwisted CNBs\cite{Wang:2024aa} or clarenes\cite{kek_clr} and, 
for each isomer, we explored every possible position around the nanoblet for the half-twist 
to generate the M{\"o}bius structures. 
In total, we have obtained 1430 isomers for [16]M{\"o}bius CNBs (12,4) 
and 486 isomers for [40]M{\"o}bius clarenes. 
Given such a large number of isomers, 
we calculated HOMO--LUMO gaps using the computationally less demanding xTB method. 
As shown in Fig.~\ref{fig:gap}, 
the HOMO--LUMO gaps predicted by the SHMO theory correlate impressively well 
with the xTB-calculated energy gaps ($R^2>0.99$ in both cases). 
From the linear least-squares fit of the data points in Fig.~\ref{fig:gap}, 
we can estimate the resonance integral, $\beta$, as the slope of the regression line. 
The obtained values are approximately $-3.03$ eV for [16]Möbius CNBs (12,4) and $-2.74$ eV for [40]Möbius clarenes, 
which are in reasonable agreement with the estimated values for CNBs. \cite{Wang:2024aa}

\section{\label{sec:conclusions}
Concluding remarks
}

In this study, we have examined the SHMO theory for M{\"o}bius CNBs. 
We first demonstrated that the adjacency matrix 
preserves its eigenvalues, with its eigenvectors only flipping their directions 
when the sign inversion position changes along the minimum half-cut of the molecular graph of the nanobelt. 
This sign inversion theorem ensures that phase inversion elements in the H{\"u}ckel Hamiltonian matrix can be set rather flexibly 
while yielding consistent SHMO results for the $\pi$ electronic structure. 
As a consequence of the theorem and the M{\"o}bius topology, 
low-lying HMOs exhibit exactly doubly degeneracy for M{\"o}bius polyenes and 
M{\"o}bius CNBs with {\em topological} symmetry, 
which we may refer to as {\em topological degeneracy}. 
For the realistic DFT-calculated $\pi$ MOs of a M{\"o}bius CNB, 
this strict topological degeneracy does not persist due to the breakdown of {\em geometric}  symmetry.  
Nevertheless, the lower $\pi$ MOs may still be doubly quasidegenerate because of topological symmetry. 
Moreover, the theorem's proof implies that for any given M{\"o}bius CNB  
successive negation operations can generate numerous isospectral adjacency matrices 
that do not necessarily correspond to minimum half-cuts. 

We then vindicated the SHMO theory for M{\"o}bius CNBs 
by comparing its results with those obtained from the more accurate DFT and xTB calculations. 
To facilitate a better comparison between the $\pi$ MOs  
from the SHMO theory (a purely $\pi$ model) and the DFT calculations 
(which typically result in CMOs of $\sigma$-$\pi$ mixed character), 
we developed an LMO-based technique to isolate $\pi$ subsystem from the whole molecule. 
By diagonalizing the transformed Fock matrix associated with the $\pi$ subsystem, 
we attain a well-defined set of $\pi$-CMOs. 
These $\pi$-CMOs exhibit impressive similarity to the HMOs in terms of orbital shape and distribution 
for the two test cases: 
the experimentally synthesized [50]M{\"o}bius CNB (25,25) 
and the hypothetical [26]M{\"o}bius clarene <2,4,6,4,2,8>. 
This technique also enables us to compute the Wiberg bond indices for $\pi$ electrons 
in M{\"o}bius systems as well as other nonplanar $\pi$-conjugated systems 
such as fullerenes, carboncones \cite{Chen:2022aa}, and CNBs.\cite{Wang:2024aa} 
The Coulson bond orders from the SHMO theory show a strong correlation 
with the Wiberg bond indices analyzed from DFT calculations 
for all C--C bonds in both test cases. 
Finally, we showed that the HOMO--LUMO gap predicted by the SHMO model 
correlates extraordinarily well with those obtained from xTB calculations 
across a large number of isomers of [16]M{\"o}bius CNBs (12,4) and of [40]M{\"o}bius clarenes. 

These results provide compelling evidence that  
the SHMO theory effectively describes the $\pi$ electronic structure and $\pi$ bonding for M{\"o}bius CNBs. 
This finding underscores that  
the $\pi$ electronic structure of M{\"o}bius CNBs is essentially determined by their topology, 
in spite of their deformed geometry where the twist is not uniformly distributed around the nanobelt. 
Additionally, the SHMO theory reveals an interesting distinction:  
the low-lying $\pi$ MOs for M{\"o}bius conjugated macrocycles have an odd number of nodal planes, 
while those for H{\"u}ckel systems display an even number. 
This characteristic can serve as a valuable criterion for distinguishing between M{\"o}bius and H{\"u}ckel systems. 

The findings from this study are fundamental for understanding the stability and 
physicochemical properties of M{\"o}bius CNBs. 
For instance, the recently synthesized [50]M{\"o}bius CNB (25,25) \cite{Segawa:2022aa} 
exhibited notable photophysical properties, including greenish-blue fluorescence emission, 
which is directly related to its $\pi$ electronic structure. 
Additionally, the chemical stability and reactivity of M{\"o}bius CNBs are primarily determined 
by their frontier MOs, which can be effectively elucidated using the SHMO model. 
Although the present study focuses on double-stranded, singly twisted M{\"o}bius CNBs, 
the results should be extendable to the thicker, multilayered\cite{Wang:2024aa} M{\"o}bius CNBs with 
triple, \cite{Fan:2023aa} quintuple or other odd numbers of twists.

\section{\label{sec:methods}
Methods
}

Using the \textsc{Gaussian} 16 \cite{g16} program, 
we carried out the DFT calculations at the B3LYP\cite{Lee88,Becke93}/6-31G* level. 
The total DFT energy includes dispersion correction 
according to Grimme's DFT-D3 empirical formula with Becke--Johnson damping. \cite{Grimme:2011aa} 
We conducted the GFN2-xTB \cite{Grimme:2017wn,Bannwarth:2019th} computations 
using the \textsc{xtb} 6.3.3 \cite{xtb-program}  program. 
We fully optimized the geometries of all considered molecules without constraints. 
We utilized the NBO 7.0 program \cite{nbo7} to obtain the NAO basis  from the original AO basis set. 
We performed Pipek--Mezey localization of MOs using our open-source software 
\textsc{EzReson} (version 3.0). \cite{ezreson,ezreson1,ezreson2,Wang:2022aa,Wang:2022tg} 
We used the Jmol (version 16.2.17) \cite{Jmol} program to depict the 3D molecular structures 
and render the MOs. 

Given that simple H{\"u}ckel method is topologically-based, 
it is valuable to visualize the HMOs on top of the 3D framework of the actual molecule, 
allowing for a direct comparison between the $\pi$ MOs  obtained from the SHMO and DFT calculations.
To this end, we first create at each of the carbon atoms an idealized 2p$_z$ atomic orbital,  
expressed in atomic units as follows:\cite{Levine_QC}
\begin{equation}
\phi_\mathrm{2p} \varpropto Z_\mathrm{eff}^{5/2} z \exp{(-Z_\mathrm{eff} r)}
\end{equation}
where $Z_\mathrm{eff}$ is the effective nuclear charge for carbon's 2p electrons  
and takes a recommended value of 3.136,\cite{Clementi:1963aa,Clementi1967AtomicSC}
consistent with Slater's rule.\cite{PhysRev.36.57}
Each 2p$_z$ orbital is centered at the position of its respective carbon atom 
and aligned according to the $\pi$-orbital axis vector (POAV) \cite{POAV,Haddon:1988aa} of that carbon atom. 
The HMOs are then constructed as linear combinations of these 2p$_z$ AOs, 
embedding the coefficients given by the simple H{\"u}ckel method. 
By calculating the volumetric data from these HMO wave functions, 
we generate \textsc{Gaussian} cube files for visualizing of H{\"u}ckel orbitals. 
This approach has been previously implemented in our open-source package 
\textsc{FullFun} \cite{FullFun,Wang2016,Wang:2017xy,Wang:2018uv} 
for the visualization of HMOs for fullerenes.

\begin{acknowledgments}
The author wish to acknowledge the financial support of 
the National Natural Science Foundation of China (22073080, 22473097) 
and the Double Innovation Talent Program of Jiangsu Province (No. JSSCRC2021542).
\end{acknowledgments}

\section*{AUTHOR DECLARATIONS}

\subsection*{Conflict of Interest}
The author has no conflicts to disclose.

\subsection*{Author Contributions}

\textbf{Yang Wang:} Conceptualization; Data Curation; Formal Analysis; Funding Acquisition;
Investigation; Methodology; Resources; Software; Validation; Visualization; 
Writing/Original Draft Preparation; Writing/Review \& Editing

\section*{Data Availability Statement}

The data that support the findings of this study are available from the corresponding author upon reasonable request.

\appendix

\section{\label{app:sigmat}
Adjacency matrix after negation operation on a given node}

We prove that the negation operation on node $p$ in a graph, $\hat{N}_p$, 
negates all elements in row $p$ and column $p$ of the adjacency matrix, $\mathbf{A}$. 

By definition, the elements of the signature matrix associated with $\hat{N}_p$ can be compactly written as
\begin{equation}\label{eq:sigmat}
(\mathbf{S}_p)_{ij} = \delta_{ij} (1 - 2\delta_{jp}), 
\end{equation}
where $\delta$ denotes the Kronecker delta. 
Using matrix multiplication, we compute the elements of the matrix product $\mathbf{S}_p \mathbf{A}$ as  
\begin{equation}
(\mathbf{S}_p \mathbf{A})_{ij} = \sum_k{ (\mathbf{S}_p)_{ik} \mathbf{A}_{kj} } 
= (1 - 2\delta_{ip}) \mathbf{A}_{ij}. 
\end{equation}
The final adjacency matrix after applying the negation operation is given by  
\begin{eqnarray}
(\hat{N}_p \mathbf{A})_{ij} & = & (\mathbf{S}_p^{-1}  \mathbf{A} \mathbf{S}_p)_{ij} 
= \sum_k{ (\mathbf{S}_p \mathbf{A})_{ik} (\mathbf{S}_p)_{kj} } \nonumber \\
& = & \sum_k{ (1 - 2\delta_{ip}) \mathbf{A}_{ik} \delta_{kj} (1 - 2\delta_{jp})} \nonumber \\
& = & (1 - 2\delta_{ip}) (1 - 2\delta_{jp}) \mathbf{A}_{ij} .
\end{eqnarray}
More explicitly, the transformed adjacency matrix can be expressed as 
\begin{equation}\label{eq:adjmat_neg}
(\hat{N}_p \mathbf{A})_{ij} = 
  \begin{cases}
    \phantom{-}0       & \quad \text{if } i = j = p\\
    -\mathbf{A}_{ij}       & \quad \text{if } i = p \neq j \text{ or } j = p \neq i\\
    \phantom{-}\mathbf{A}_{ij}  & \quad \text{otherwise.}
  \end{cases}
\end{equation}
Thus, all elements in row $p$ and column $p$ are negated, 
while all other entries remain unchanged from the original adjacency matrix.

\section{\label{app:cuts}
Minimum half-cuts for M{\"o}bius CNBs}

\begin{figure}[h]
\includegraphics[width=0.5\textwidth]{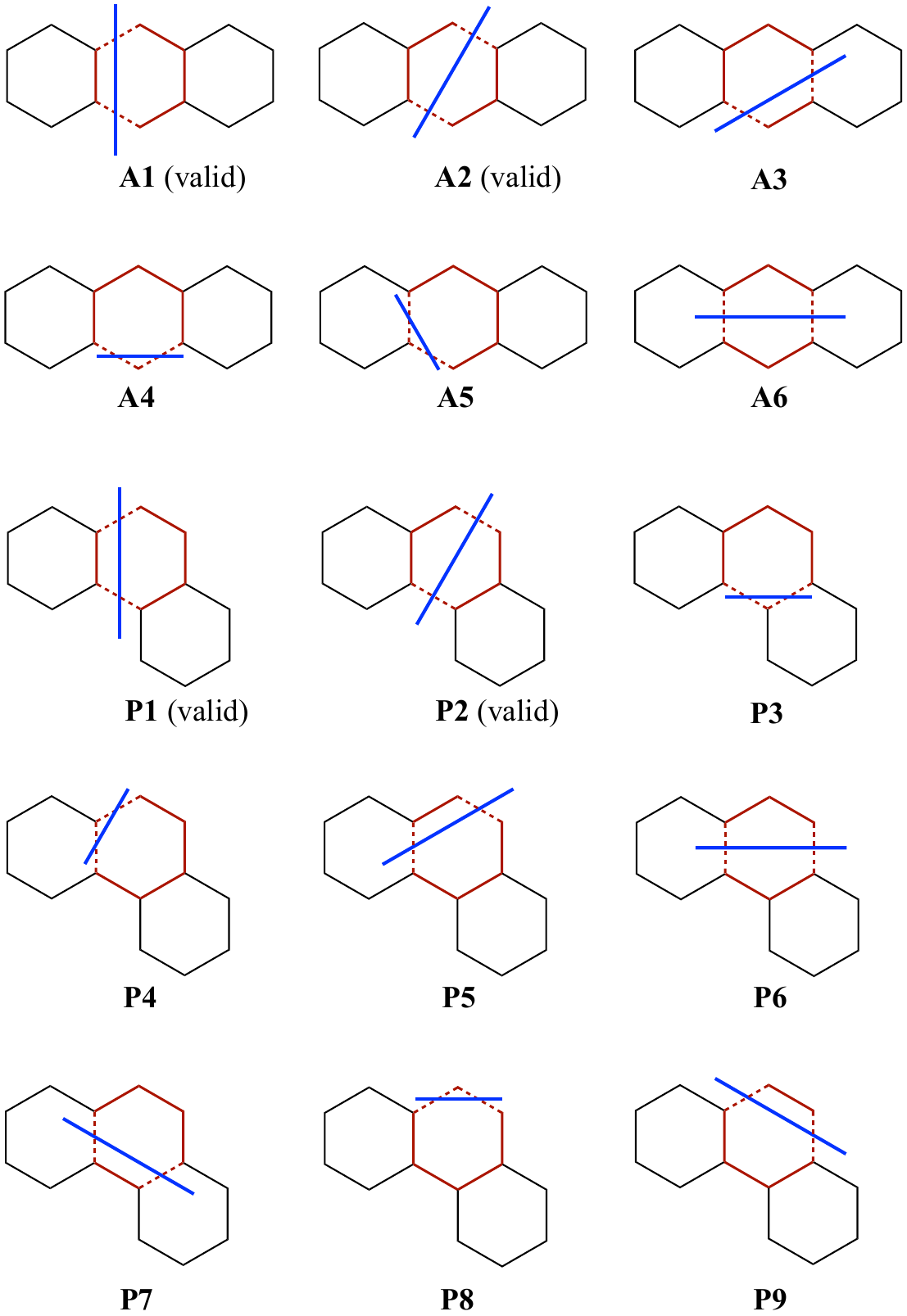}
\caption{\label{fig:patterns_cut} 
All possible topologically nonequivalent ways of breaking two C--C bonds within a benzenoid ring 
in any M{\"o}bius CNB. 
\textbf{A1}--\textbf{A6} are the six possible bond disconnection patterns for anthracene-type central rings, 
while \textbf{P1}--\textbf{P9} are the nine possible scenarios for phenanthrene-type central rings.
The central rings are highlighted in brown with dashed lines denoting broken C--C bonds. 
The blue line drawn in each pattern indicates the bond cleavage. 
}
\end{figure}

A minimum half-cut in the molecular graph of a M{\"o}bius CNB involves removing two edges. 
In other words, two C--C bonds need to be disconnected to open the macrocycle of the double-stranded carbon framework of a M{\"o}bius CNB.\cite{Segawa:2016aa,Imoto:2023aa} 
Since the macrocycle remains closed if the two broken bonds are from different rings, 
it is evident that both broken bonds must be within the same ring. 
In double-stranded M{\"o}bius CNBs, the benzenoid rings fall into two annulation types: 
anthracene and phenanthrene types.
An anthracene-type ring is aligned in a straight row with its two neighboring rings, 
similar to the ring arrangement in an anthracene molecule. 
In comparison, a phenanthrene-type ring connects with its two neighboring rings in an angled manner. 
We can enumerate all possible scenarios for removing two edges within the same benzenoid ring 
by considering both annulation types. 
As shown in Fig.~\ref{fig:patterns_cut}, 
there are six topologically nonequivalent possibilities for an anthracene-type ring (\textbf{A1}--\textbf{A6}), 
but only \textbf{A1} and \textbf{A2} open the molecular framework. 
For a phenanthrene-type ring, 
among the nine nonequivalent possibilities (\textbf{P1}--\textbf{P9} in Fig.~\ref{fig:patterns_cut}), 
only \textbf{P1} and \textbf{P2} are valid ways to open the macrocycle. 
Therefore, there are only four possible topologically nonequivalent minimum half-cuts for M{\"o}bius CNBs: 
\textbf{A1}, \textbf{A2}, \textbf{P1}, and \textbf{P2}.

\section*{References}
%

\end{document}